\documentclass[graybox]{svmult}

\usepackage{mathptmx}       
\usepackage{helvet}         
\usepackage{courier}        
\usepackage{type1cm}        
\usepackage{makeidx}         
\usepackage{graphicx}        
\usepackage{multicol}        
\usepackage[bottom]{footmisc}

\usepackage{booktabs}

\makeindex             

 \def\sig#1#2{}
 \def\F{_{\sss F}}

 \newenvironment{fanc}
 {\vskip 0.5 cm\rule{1ex}{1ex}\hspace{\stretch{1}}\noindent}
 {\hspace{\stretch{1}}\rule{1ex}{1ex}\vskip 0.5cm}

\def\bay{\begin{array}}
\def\eay{\end{array}}
\def\bit{\begin{itemize}}
\def\beit{\begin{itemize}}
\def\eit{\end{itemize}}

\def\e{_{\rm e}}

\def\apx{^\text{apx}}

\def\intr{ {\int d^3 r\,}}

\def\n{\rho}

\def\sss{\scriptscriptstyle\rm}

\def\dn{_\downarrow}

\def\g{_\gamma}

\def\l{^\lambda}

\def\1var{(\bx_1...\bx\N)}

\def\x{_{\sss X}}
\def\c{_{\sss C}}
\def\s{_{\sss S}}
\def\xc{_{\sss XC}}

\def\Hxc{_{\sss HXC}}

\def\N{_{\sss N}}
\def\H{_{\sss H}}

\def\LDA{^{\rm LDA}}

\def\unif{^{\rm unif}}

\def\ee{_{\rm ee}}

\def\sph_int{ {\int d^3 r}}

\def\intr{\int d^3r\,}
\def\intrp{\int d^3r'\,}

\def\br{{\bf r}}

\def\bx{{x}}

\def\bea{\begin{eqnarray}}
\def\eea{\end{eqnarray}}
\def\ben{\begin{equation}}
\def\een{\end{equation}}
\def\benu{\begin{enumerate}}
\def\enu{\end{enumerate}}

\def\bei{\begin{itemize}}
\def\eei{\end{itemize}}
\def\beit{\begin{itemize}}
\def\eit{\end{itemize}}
\def\benu{\begin{enumerate}}
\def\enu{\end{enumerate}}

\def\half{\frac{1}{2}}

\usepackage{lastpage}

\usepackage{fancyhdr}
\definecolor{SECOL}{rgb}{0.1,0.2,0.7} 

\fancypagestyle{plain}{%
  \fancyhf{}%
}

\let\textcite\relax
\makeatletter
\DeclareRobustCommand{\MakeUppercase}[1]{{%
      \def\i{I}\def\j{J}%
      \def\reserved@a##1##2{\let##1##2\reserved@a}%
      \expandafter\reserved@a\@uclclist\reserved@b{\reserved@b\@gobble}%
      \protected@edef\reserved@a{\uppercase{#1}}%
      \reserved@a
   }}   
\DeclareRobustCommand{\MakeLowercase}[1]{{%
      \def\reserved@a##1##2{\let##2##1\reserved@a}%
      \expandafter\reserved@a\@uclclist\reserved@b{\reserved@b\@gobble}%
      \protected@edef\reserved@a{\lowercase{#1}}%
      \reserved@a
   }}   
\makeatother
\expandafter\let\csname ver@natbib.sty\endcsname\relax

\usepackage[style=alphabetic,backend=bibtex]{biblatex}
\DeclareFieldFormat{labelalpha}{\thefield{entrykey}}
\DeclareFieldFormat{extraalpha}{}

\renewcommand{\bibliography}[1]{}

\begin{document}

\title*{Warming Up Density Functional Theory}
\author{Justin C. Smith, Francisca Sagredo, and Kieron Burke}
\institute{
Justin C. Smith \at Department of Physics and Astronomy, University of California, Irvine, CA 92697, \email{justincs@uci.edu}
\and Francisca Sagredo \at Department of Chemistry, University of California, Irvine, CA 92697, \email{fsagredo@uci.edu}
\and Kieron Burke \at Department of Chemistry, University of California, Irvine, CA 92697 \email{kieron@uci.edu}
}

%
%
\maketitle

\abstract*{
Density functional theory (DFT) has become the most popular approach
to electronic structure across disciplines, especially in material
and chemical sciences.  Last year, at least 30,000 papers used DFT
to make useful predictions or give insight into an enormous diversity
of scientific problems, ranging from battery development to solar
cell efficiency and far beyond.  The success of this field has been
driven by usefully accurate approximations based on known exact
conditions and careful testing and validation. 
In the last decade, applications of DFT in a new area, warm dense matter, have exploded.
DFT is revolutionizing simulations of warm dense matter
including applications in controlled fusion, planetary interiors, and other areas
of high energy density physics.
Over the past decade or so, molecular
dynamics calculations driven by modern density
functional theory have played a crucial role in bringing chemical
realism to these applications, often (but not always) with excellent
agreement with experiment.
This chapter summarizes recent work from our group on density
functional theory at non-zero temperatures, which we call thermal DFT.   
We explain the relevance of this work in the context of warm dense
matter, and the importance of quantum chemistry to this regime.
We illustrate many basic concepts on a simple model system, the
asymmetric Hubbard dimer.
}

\abstract{
Density functional theory (DFT) has become the most popular approach
to electronic structure across disciplines, especially in material
and chemical sciences.  Last year, at least 30,000 papers used DFT
to make useful predictions or give insight into an enormous diversity
of scientific problems, ranging from battery development to solar
cell efficiency and far beyond.  The success of this field has been
driven by usefully accurate approximations based on known exact
conditions and careful testing and validation. 
In the last decade, applications of DFT in a new area, warm dense matter, have exploded.
DFT is revolutionizing simulations of warm dense matter
including applications in controlled fusion, planetary interiors, and other areas
of high energy density physics.
Over the past decade or so, molecular
dynamics calculations driven by modern density
functional theory have played a crucial role in bringing chemical
realism to these applications, often (but not always) with excellent
agreement with experiment.
This chapter summarizes recent work from our group on density
functional theory at non-zero temperatures, which we call thermal DFT.   
We explain the relevance of this work in the context of warm dense
matter, and the importance of quantum chemistry to this regime.
We illustrate many basic concepts on a simple model system, the
asymmetric Hubbard dimer.
}

\section{Introduction}
\label{intro}

{\bf Warm dense matter:}
The study of warm dense matter (WDM) is a rapidly growing multidisciplinary field that spans
many branches of physics, including for example
astrophysics, geophysics, and attosecond physics\cite{MD06,DOE09,LHR09,KDBL15,KD09,KDP15,HRD08,KRDM08,RMCH10,SEJD14,GDRT14}.
Classical (or semiclassical)
plasma physics is accurate for sufficiently high temperatures and sufficiently
diffuse matter\cite{I04}.  The name WDM implies too cool and too dense for such methods
to be accurate, and this regime has often been referred to as the malfunction junction,
because of its difficulty\cite{DOE09}.
Many excellent schemes have been developed over the decades within plasma
physics for dealing with the variety of equilibrium and non-equilibrium
phenomena accessed by both people and nature under the relevant conditions\cite{BL04}.
These include DFT at the Thomas-Fermi level (for very
high temperatures) and use of the local density approximation (LDA)
within Kohn-Sham (KS) DFT at cold to moderate temperatures (at very
high temperatures, sums over unoccupied orbitals fail to converge).
The LDA can include thermal XC corrections based on those of the uniform
gas, for which simple parametrizations have long existed\cite{SD13b,KSDT14}.

{\bf Electronic structure theory:}
On the other hand, condensed matter physicists, quantum chemists, and computational
materials scientists have an enormously well-developed suite of methods for
performing electronic structure calculations at temperatures at which the electrons
are essentially in their ground-state (GS), say, 10,000K or less\cite{B12}.  The starting point
of many (but not all) such calculations is the KS method of DFT
for treating the electrons\cite{KS65}.  Almost all such calculations are within
the Born-Oppenheimer approximation, and ab initio molecular dynamics (AIMD) is
a standard technique, in which KS-DFT is used for the electronic structure, while
Newton's equations are solved for the ions\cite{CP85}.  

{\bf DFT in WDM:}
In the last decade or so, standard methods from the electronic structure of materials
have had an enormous impact in warm dense matter, where AIMD is often called QMD,
quantum molecular dynamics\cite{GDRT14}.  Typically a standard code such as VASP is run
to perform MD\cite{KRDM08}.  
In WDM, the temperatures are a noticeable fraction of the Fermi energy, and thus the generalization of DFT to thermal systems must be used. Such simulations are 
computationally demanding 
but they have the crucial feature of including
realistic chemical structure, which is difficult to include with any other method while
remaining computationally feasible.
Moreover, they are in principle exact\cite{M65,KS65}, if the exact temperature-dependent
exchange-correlation free energy could be used because of Mermin's theorem establishing thermal DFT(thDFT).  
In practice, some standard ground-state approximation is usually used.
(There are also quantum Monte Carlo calculations which are typically even more
computationally expensive\cite{MD00,FBEF01,M09b,SBFH11,DM12,SGVB15,DGSM16}.
The beauty of the QMD approach is that it can provide chemically realistic simulations
at costs that make useful applications accessible\cite{MMPC12}.) 
 There have been many successes,
such as simulation of Hugoniot curves measured by the $Z$ machine\cite{RMCH10} or 
a new phase diagram for 
high density water which resulted in
improved predictions for the structure of Neptune\cite{MD06}.
Because of these successes, QMD has rapidly
become a standard technique in this field. 

{\bf Missing temperature dependence:}
However, the reliability and domain of applicability of QMD calculations are even less well
understood than in GS simulations.   At the equilibrium level of calculation,
vital for equations of state under WDM conditions and the calculation of
free-energy curves, a standard generalized gradient
approximation (GGA) calculation using, e.g., PBE\cite{PBE96}, is often (but
not always) deemed sufficient, just as it is for many GS materials properties.
Such a calculation ignores thermal exchange-correlation (XC) corrections, i.e., the changes
in XC as the temperature increases, which are related to 
entropic effects.  We believe we know these well for a uniform gas (although
see the recent string of QMC papers\cite{SGVB15,DGSM16} and 
parametrizations\cite{KSDT14}), but such corrections will be unbalanced if applied to a GGA such
as PBE.  So how big a problem is the neglect of such corrections?

{\bf (A little) beyond equilibrium:}
On the other hand, many experimental probes of WDM extract response functions
such as electrical or thermal conductivity\cite{MD06}.  These are always calculated from
the equilibrium KS orbitals, albeit at finite temperature.  Work on 
molecular electronics  shows that such evaluations
suffer both from inaccuracies in the positions of KS orbitals due
to deficiencies in XC approximations, and also require 
further XC corrections, even if the {\em exact} equilibrium XC functional
were used\cite{TFSB05,QVCL07,KCBC08}.

\begin{table}[htb]
\begin{tabular}{|l | l| l | l |}
\toprule
\hline
Acronym & Meaning & Acronym & Meaning\\
\hline
GGA & Generalized Gradient Approx. &RPA & Random Phase Approximation \\
GS & ground-state & TDDFT & Time-dependent DFT \\ 
HXC & Hartree XC & thDFT & thermal DFT\\ 
KS & Kohn-Sham & unif & uniform gas \\
LDA & Local Density Approx. & XC & exchange-correlation\\
PBE & Perdew-Burke-Ernzerhof &  ZTA & Zero-Temperature Approx.  \\
QMC & quantum Monte Carlo & & \\
\hline
\bottomrule
\end{tabular}
\label{acrodef}
\caption{Acronyms frequently used in this chapter.}
\end{table}

\def\F{_{\sss F}}
\def\t{^{\tau}}
\def\unift{^{{\rm unif},\tau}}
\def\intr{\int d^3r\,}
\def\intrp{\int d^3r'\,}
\def\e{_\text{e}}
\def\apx{^\text{apx}}
\def\apxo{^\text{apx,0}}
\def\tp{^{\tau'}}
\def\ttp{^{\tau,\tau'}}
\def\inft{\tau''}
\def\wt{w}
\def\wtb{\overline{\wt}}
\def\FF{A}
\def\Gam{\hat{\Gamma}}
\def\st{_{\sqrt{\tau'/\tau}}}
\def\g{_\gamma}
\def\l{^\lambda}

\section{Background}

{\bf Generalities:}
Everything described within uses atomic units, is non-relativistic
and does not include external magnetic fields.  Unless otherwise
noted, all results are for the electronic contributions
within the Born-Oppenheimer approximation.  While all results are
stated for density functionals, in practice, they are always generalized to
spin-density functionals in the usual way.

\subsection{Ground-state DFT}

{\bf Hohenberg-Kohn functional:}
Just over 50 years ago, in 1964, Hohenberg and Kohn  wrote down the foundations of modern DFT\cite{HK64}.
They start with the many-body Hamiltonian
\ben
\hat{H} = \hat{T} + \hat{V}\ee + \hat{V},
\een
where $\hat{T}$, $\hat{V}\ee$, and $\hat{V}$ are the kinetic, electron-electron, and potential energy operators, respectively. 
Assuming a non-degenerate ground-state, they proved by \emph{reductio ad absurdum} that the
external potential, $v(\br)$ is a unique functional of the density $\n(\br)$, and therefore
all observables are also density functionals.  More directly
Levy defines the functional
\ben
F[\n] = \min_{\Psi\to\n} \langle \Psi | \hat{T} + \hat{V}\ee | \Psi \rangle,
\label{Ffun}
\een
where $\Psi$ is normalized and antisymmetric, and uses it to define the energy functional
\ben
E_v[\n] = F[\n] + \intr v(\br) \n(\br),
\een
whose minimization over normalized non-negative densities with finite kinetic energy
yields the ground-state energy and density\cite{L81}.

{\bf Kohn-Sham scheme:}
In 1965, Mermin generalized the Hohenberg-Kohn theorems 
for electrons in the grand canonical potential with fixed non-zero temperature $\tau$
and chemical potential $\mu$\cite{M65}.
Later in 1965, Kohn and Sham created an exact method to construct the universal functional (see Eq. (\ref{FfunKS})).
The Kohn-Sham scheme imagines a system of $N$ non-interacting electrons that yield
the electronic density of the original interacting $N$ electron system.  These fictitious electrons 
sit in a new external potential called the KS potential. The KS scheme is written
as a set of equations that must be solved self-consistently:
\ben
\left\{-\half \nabla^2 + v\s(\br) \right\}\phi_i(\br) = \epsilon_i \phi_i(\br), ~~~ \n(\br) = \sum_i^N |\phi_i(\br)|^2,
\label{KSeq}
\een
\ben
v\s(\br) = v(\br) + v\H(\br) + v\xc(\br),~~~
v\xc(\br) = \frac{\delta E\xc[\n]}{\delta \n(\br)},
\label{XCpot}
\een
where $\phi_i(\br)$ and $\epsilon_i$ are the KS orbitals and energies, $v\H(\br)$ is the classical
Hartree potential, and $v\xc(\br)$ is the exchange-correlation potential defined
by the unknown XC energy, $E\xc$, in Eq. (\ref{XCpot}). 
These must be solved self-consistently
since the Hartree potential and $E\xc$ depend explicitly on the density. 
Lastly, the total energy can be found via
\ben
F[\n] = T\s[\n] + U\H[\n] + E\xc[\n]
\label{FfunKS}
\een
where $T\s$ is the kinetic energy of the KS electrons and $U\H$ is the Hartree energy.

In practice, an approximation
to $E\xc$ must be supplied.
There exists a wealth of approximations for $E\xc$\cite{MOB12}. 
The simplest, LDA, uses the XC per electron of the homogeneous electron gas\cite{PW92}:
\ben
E\xc\LDA[\n] = \intr e\xc\unif(\n(\br)) 
\label{LDAdef} 
\een
where $e\xc\unif$ is the XC energy density of a uniform gas with density $\n(\br)$.
We can imagine going up a ladder by adding in more
ingredients (like gradients of the density\cite{PBE96}) and obeying different or additional conditions to make more complicated and more accurate functionals\cite{PRCS09}. 
For the exact $E\xc$, these equations have been proven to converge\cite{WSBW13}.

\subsection{Asymmetric Hubbard dimer and its relevance}

\def\dv{\Delta v}
\def\dn{\Delta n}

Throughout this chapter we illustrate results with the simplest interesting model of an 
interacting system. 
This model is the  asymmetric Hubbard dimer\cite{CFSB15}. The Hubbard dimer crosses
the divide between the weakly and strongly correlated communities.
Previous work has shown that the basic concepts of ground-state KS-DFT apply to this model and here we
demonstrate similar applicability to thermal DFT.
The Hamiltonian is given by
\begin{equation}
\hat{H} = -t \sum_\sigma (\hat{c}_{1\sigma}^\dagger \hat{c}_{2\sigma} + H.c.) +
\sum_{i=1}^2(U \hat{n}_{i\uparrow} \hat{n}_{i\downarrow} + v_i \hat{n}_i)
\label{HubbHam}
\end{equation}
where $\hat{c}_{i\sigma}^\dagger (\hat{c}_{i\sigma})$ are electron creation (annihilation) operators
and $\hat{n}_{i\sigma} = \hat{c}_{i\sigma}^\dagger\hat{c}_{i\sigma}$ are number operators.
$t$ is the strength of electron hopping between sites, $U$ is the ultra-short range Coulomb 
repulsion for when both electrons are on the same site, and $v_i$ is the on-site external potential. We choose, without loss of generality, $v_1 +v_2 =0$ then $\Delta v = v_2 - v_1$ and the occupation difference, the analog of density, is $\Delta n = n_2 - n_1$.

\begin{figure}[htb]
\centering
\includegraphics[width=.8\textwidth]{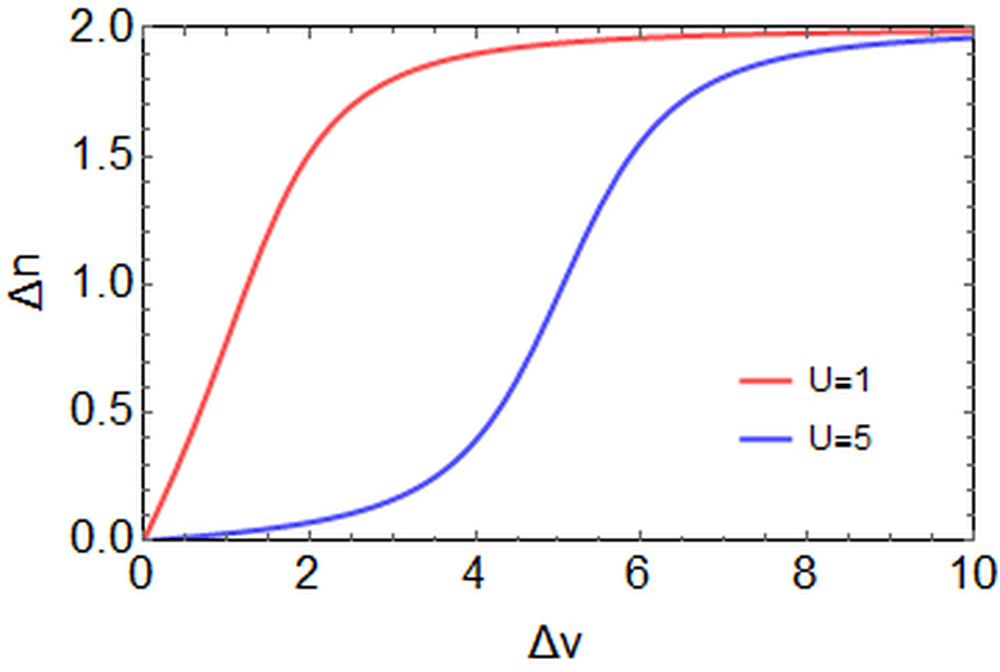}
\caption{
On-site occupations as a function of on-site potential difference for $U=1$
and 5 in the asymmetric Hubbard dimer.  The HK theorem guarantees that each
function is invertible.  There is a simple analytic result for $U=0$, and
for large $U$, the relation tends toward a (smoothed) step function, with the
step at $\dv=U$.
}
\label{hubfigdnGS}
\end{figure}

In Fig. \ref{hubfigdnGS} we plot the density $\dn$ versus asymmetry $\dv$ in the Hubbard dimer.
The Hohenberg-Kohn theorem applies to this Hamiltonian, and guarantees $\dn(\dv)$ is an
invertible function for any value of $U$.
The main physics is a competition between asymmetry and interaction strength. The weakly
correlated regime is $U < \Delta v$ and the opposite is strong correlation.
Increasing $\dv$ pushes the electrons onto a single site, thus $\dn$ approaches 2. Likewise,
for small $\dv$ or large $U$ the electrons are apart and $\dn$ tends to 0. This is made most 
clear by the extreme cases, i.e.,
\ben
\label{dnU}
|\dn_{U=0}(\dv)| = 2\dv/\sqrt{(2\,t)^2 + \dv^2}, ~~~~~~~
|\dn_{U\to \infty}(\dv)| 
\to 2\theta(\dv-U).
\een
The ability to vary $U$ and move continuously from weak to strong correlation in a model
that is analytically solvable makes the Hubbard dimer an excellent illustrator of how
KS-DFT works\cite{CFSB15}.

\subsection{Ensemble DFT as a route to excitation energies}

In this section we take a quick aside to overview ensemble DFT (eDFT), a close cousin of
thermal DFT.

{\bf Excitations in DFT:}
Although time-dependent DFT (TDDFT) is the standard method used to determine the excited states of a system \cite{M16}, there are still many deficiencies, due to crude approximations to the XC functional as well as   being unable to approximate multiple excitations, charge transfer excitations, canonical intersections, and polarizabilities of long-chain polymers; all things that can be
important for photochemistry\cite{MZCB04,DWH03,T03,LKQM06,FBLB02}. 
Ensemble DFT is a time-independent alternative to the standard TDDFT that can be a useful method for extracting excited states. Naturally, since eDFT and TDDFT are based on two different fundamental theories, it is possible to use eDFT on different systems to those of the traditional method and expect different successes and likewise different failures.  

\def\bw{{\bf w}}
{\bf Ensemble variational principle:}
eDFT is based on a variational principle made up of ensembles of ground and excited states
\cite{T79}.
These ensembles are made of decreasing weights, with the ground state always having the highest weight. 
\ben
E^{\bw} \leq  \sum_{k=0}^{M-1}  w_{k} \langle \Psi_{k}| \hat{H} | \Psi_{k}\rangle,~~~~
w_{0} \geq w_{1} \geq ... \geq w_{k} \geq 0
\label{eDFT1eq}
\een
where all $\Psi_k$ are normalized, antisymmetric, and mutually orthogonal, 
$\bw=(w_{0}, w_{1}, ... w_{k} )$, and the sum of all weights is 1.
The ensemble-weighted density is
\ben
\n^{\bw}(\br) =  \sum_{k=0}^{M-1}  w_{k} \n_{\Psi_{k}}(\br).
\label{eDFT2}
\een
Just as in the ground-state case, a one-to-one correspondence from the weighted
density to the potential can be established \cite{GOK88}, and applying this
to a non-interacting system of the same weighted density can
be used to construct a KS eDFT. 
From this KS system it is in principle possible to extract the
exact excited states of the system.

{\bf Relation to thermal DFT:}
The connection to thermal DFT is natural and straight forward. Thermal DFT is
a special case of eDFT.  In thDFT,
one chooses the ensemble to be the grand canonical ensemble with the
usual Boltzmann factors for the weights.   However, unlike eDFT, the weights
themselves depend on the eigenvalues of the Hamiltonian, including the strength
of the interaction.  Thus the weights in the KS system are different from those
of the interacting system.  In most applications of eDFT, the weights are chosen
to be the same in both the physical and non-interacting systems.

{\bf History:}
Ensemble DFT was originally proposed by Gross-Olivera-Kohn in 1988\cite{GOK88}, 
but, like thermal DFT, there has been slow progress over the last 30 years due to
a lack of useful approximations to the XC functional. Many of these
difficulties arise from the so-called ghost interaction errors that
occur in the Hartree energy\cite{GPG96}. More specifically these
ghost-interaction errors appear when only using the ground state definition
of the Hartree energy, which causes unphysical contributions and must
therefore be accounted for by using a more accurate definition
of the Hartree energy for ensembles\cite{PYTB14,YTPB14}.  

{\bf Recent progress:}
More recently, work has been done to extract the weight dependence of the KS
eigenvalues, which are required in order to extract accurate transition
frequencies\cite{PYTB14,YTPB14}. It was also found that a large
cancellation of the weight-dependence occurs in the exact ensembles. Further,
a new numerical method for inverting ensemble densities was derived for spherically
symmetric systems, and this method was also tested for cylindrically symmetric
systems. This inversion of densities to extract potentials provides a useful 
test of eDFT approximations.

Recent work combines linear interpolation with an extrapolation method in eDFT
to extract excited states that are independent of ensemble weights\cite{SHMK16}.
Also, an exact analytical expression for the exchange energy was derived, and
a generalized adiabatic connection for ensembles (GACE) was used 
to connect the ensemble XC functional with the ground state functional\cite{DMF16}. 
Finally, a ghost interaction correction has been developed for range-separated eDFT\cite{AKF16}.

{\bf A simple exercise:}
Here we show an example of the importance of the weight-dependence of
functionals in eDFT, in a seemingly simple system.  We put only one electron in the
Hubbard dimer, so there is no interaction, and its a simple tight-binding Hamiltonian.
We will derive the exact kinetic energy functional, which is an example of the non-interacting
KS kinetic energy functional.  We will also approximate it, as if we were interested
in orbital-free eDFT.

There are only two levels, the ground state and a first excited state.
Thus we can make only a bi-ensemble.
The ensemble-weighted ground-state density is
\ben
\label{dnw}
\dn^{w}= (1-w) \dn_{0} +w \dn_{1} = (1-2 w) \dn_{0}(x),
\een
where $\dn_{0}$, and $\dn_{1}$ correspond to the occupational difference
of the two sites for the ground and first-excited state, respectively, 
and $\dn_{0}(x)$ is given by Eq. (\ref{dnU}) with $U=0$.
This last result is true only because $\dn_{1}=-\dn_0$ in this simple model.
The weight as previously stated is $w \leq 0.5$.  
Similarly, the kinetic energy for a single particle in the ground-state is known,
and $T_{\sss S,1}=-T_{\sss S,0}$, so
\ben
\label{eq:Tswn}
T\s^{w}= (1-w)T_{\sss S, 0} + w T_{\sss S,1}= (1-2w) T_{\sss S,0}= (1-2w)\frac{-1}{2{\sqrt{1+x^2}}}
\een
where $x=\dv/(2\,t)$.  Using $V^w= \dv\dn^w/2$ and adding it to $T\s^w$ yields
the ensemble energy, which is exactly linear in $w$, and passes through $E_0$
at $w=0$ and (would pass through) $E_1$ at $w=1$.

This simple linearity with $w$ is true by construction of the ensemble, when energies
are plotted against $w$ for a fixed potential.  But now we show that things get
complicated when we consider them as density functionals.  Inverting the relation
between potentials and densities we find 
\ben
\label{xw}
x= \frac{ \dn^{w}}{\sqrt{(-\dn^{w})^2 + (1-2w)^2}}
\een
and inserting this into the kinetic energy yields
\ben
\label{eq:Tswn2}
T\s^{w}[\dn]= {\sqrt{-(\dn)^{2} +(1-2w)^2}}/ {2}
\een
Even in this trivial case, the
the $w$-dependence of the kinetic energy density functional is non-linear.

\begin{figure}[htb]
\centering
\includegraphics[width=.8\textwidth]{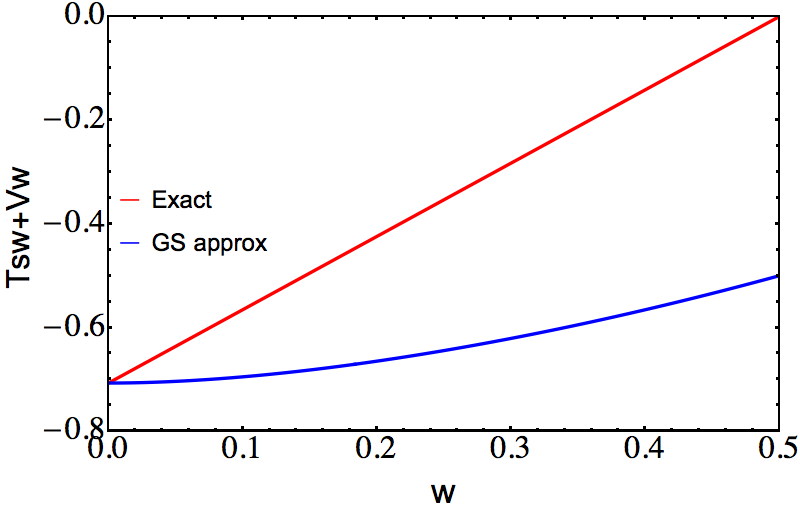}
\caption{The exact energy curve for a single particle in a Hubbard dimer 
$E_{exact}= T\s^{w}[\dn^{w}] +V^{w}$ (red), when $x=1$, in units of $2\,t$.
The blue line is the approximate energy when $T\s^w$ is replaced by its ground-state 
analog, $T\s^0$.
Notice that $w=0$ corresponds to $E_0$, which is $1/{\sqrt{2}}$ in these units.
}
\label{ensemble2}
\end{figure}

Next, we make the most naive approximation, namely to replace $T\s^w[\dn]$
with its ground-state counterpart.  This yields an approximate eDFT theory from 
which, in principle, we can estimate the energy of the first excited state.
To do this, we insert the exact $w$-dependent density of Eq. (\ref{dnw}) into 
the ground-state functional, add the exact $V^w$, and plot the resulting energy.
The exact and approximate results are shown in Fig. \ref{ensemble2}.
The approximation is very bad, yielding an excited state energy of about $-0.3$ 
instead of $0.71$, by using the value at $w=1/2$.  But it at least illustrates
the difficulties of capturing an accurate $w$-dependence in an ensemble density functional.

\subsection{Thermal DFT in a nutshell}

{\bf Mermin-Kohn-Sham equations:}
In a thermal system, Eq. (\ref{KSeq}) and (\ref{XCpot}) are generalized such that the density
 and XC potential become $\n^\tau(\br)$ and $v\xc^\tau(\br)$, 
i.e. temperature dependent, and $E\xc[\n]$ in Eq. (\ref{XCpot}) becomes $A\xc^\tau[\n]$, the exchange-correlation \emph{free} energy density functional. The density becomes
\ben
\n^\tau(\br) = \sum_i f_i |\phi^\tau_i(\br)|^2
\een
where the sum is now over all states and $f_i = (1 + e^{(\epsilon^\tau_i -\mu)/\tau})^{-1}$, the Fermi occupation factors. 
One of the core difficulties in thermal DFT calculations is this sum, since a huge number of
states are required once the temperature is sufficiently high. This leads to large computational
demands and convergence issues.
We call these the Mermin-Kohn-Sham equations.

To extract the total free energy from the MKS equations, we write
\ben
A^\tau[\n] = A\s^\tau[\n] - U\H[\n] + A\xc^\tau[\n] - \intr \n(\br)v\xc^\tau[\n](\br),
\een
where the MKS free energy is
\ben
A\s^\tau[\n] = \sum_i \epsilon_i^\tau[\n] - \tau S^\tau\s[\n],~~~
S^\tau\s[\n] = - \sum_i\left[f_i \log(f_i) + (1-f_i) \log(1-f_i)\right],
\een
and $S\s$ is the MKS entropy.

{\bf Exchange-correlation free energy:}
Compared to ground-state DFT, relatively few approximations have been developed
for $A\xc\t$. In active use are two
approximations: thermal LDA (thLDA) and the Zero-Temperature Approximation (ZTA). 
The former uses the temperature-dependent XC free energy of the uniform gas instead of
its ground-state analog in Eq. (\ref{LDAdef}).  The ZTA means simply using any ground-state
XC functional  instead of a temperature-dependent one.  We denote use of the exact ground-state
XC functional as exact ZTA (EZTA).
\begin{figure}[htb]
\centering
\includegraphics[width=.8\textwidth]{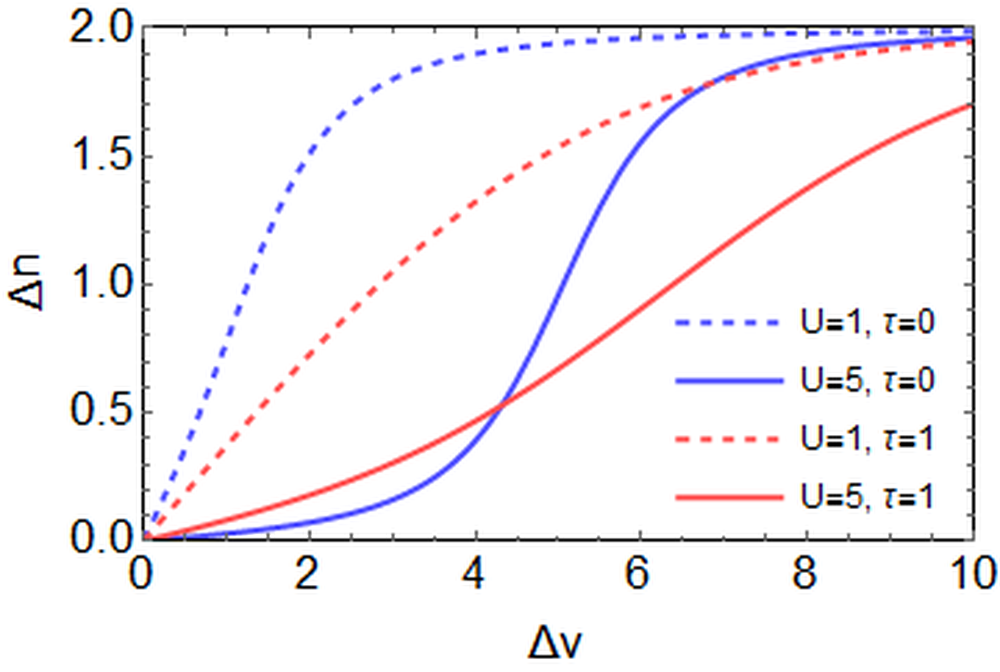}
\caption{
Effects of temperature on the on-site occupations as a function of on-site potential difference for $U=1$
and 5 in the asymmetric Hubbard dimer (see Sec. \ref{simp}) with and without temperature.
Increasing temperature pushes the electrons towards opposite
sites and lowers $\Delta n$ while increasing $\Delta v$  pushes electrons to the same site and raises $\Delta n$.
}
\label{hubfigdn}
\end{figure}

In Fig. \ref{hubfigdn} we demonstrate the effects of turning on temperature for the Hubbard 
dimer (see Sec. \ref{simp} for more information). Even a small increase in temperature can
have a big impact on a system regardless of the strength of correlation. 

\section{Some recent developments in thermal DFT}

The formalism for thermal DFT was originally developed alongside that of ground-state DFT, but in the 
intervening decades ground-state DFT (and TDDFT, for that matter) has received
significantly more attention and consequently more developments\cite{WGB05,B12}. However, In the past couple decades
thermal DFT has seen more use, and with that much more development in the past ten years.
In this section we outline some recent developments to thermal DFT from our group.

\subsection{Exact conditions and their relevance}

{\bf Zero temperature:}
One of the most crucial steps in understanding and developing functionals beyond LDA is exact conditions.
These conditions take many forms with some common examples being coordinate and
interaction scaling conditions\cite{LP85}, and bounds on the XC energy\cite{LO81}.
There are well over a dozen conditions in ground-state DFT (a recent meta-GGA functional even uses seventeen\cite{SRP15}!), but the use of exact conditions is much more nascent in thermal DFT.

\subsubsection{Coordinate-temperature scaling and the thermal connection formula}
\label{scale}

{\bf Uniform coordinate scaling:}
The most straightforward application of exact conditions to thermal DFT is by uniform
scaling of the density\cite{LP85}.  
The very basic conditions that this procedure generates in ground-state DFT
are built in to almost all modern approximations.  In a sense, this is simply dimensional
analysis, but while keeping the density fixed (which is the tricky bit).

Early work on exact conditions for thermal DFT\cite{PPFS11, PPGB13}
derived basic conditions
such as the signs of correlation quantities, including the separation into kinetic
and potential contributions, and the adiabatic connection formula at finite temperature.
More conditions come from coordinate scaling of the density, showing that is intimately
related to temperature dependence.
Examples of a few of these conditions are
\ben
F\s^\tau[\n] = \gamma^2 F\s^{\tau/\gamma^2}[\n_{1/\gamma}],~~~
S\s^\tau[\n] = S\s^{\tau/\gamma^2}[\n_{1/\gamma}],~~~
A\x^{\tau}[\n] = \gamma A\x^{\tau/\gamma^2}[\n_{1/\gamma}],
\een
where $\n\g(\br)=\gamma^3\n(\gamma\br)$.  
For any of these functionals, this means that, if you know the functional
at {\em any} one finite temperature, the
functional at {\em all} possible temperatures is available via scaling of the density.

{\bf New formulas:}
In recent work, many new formulas relating correlation components of the energy
to one another were derived\cite{PB16}, such as
\ben
K\c^{\tau,\lambda}[\n] = A\c^{\tau,\lambda}[\n] - \lambda \frac{d A\c^{\tau,\lambda}[\n]}{d\lambda},
\een
where $K\c^{\tau,\lambda}[n] = T\c^{\tau,\lambda}[\n] - \tau S\c^{\tau,\lambda}[\n]$ is the
correlation kentropy.
There was also a rewriting of the adiabatic connection
formula\cite{PPFS11,LP75}, using the relation to scaling mentioned above, yielding
the XC free
energy at temperature $\tau$:
\ben
\FF\xc\t[\n]=\frac{\tau}{2}\lim_{\inft\to\infty}
\int_\tau^{\inft}\frac{d\tau'}{\tau'^2}\, U\xc^{\tau'}[n\st],~~~~~
\n_\gamma(\br)=\gamma^3\, \n(\gamma\br),
\label{Axcth}
\een
where $U\xc\t[\n]$ is the purely potential contribution to the XC
free energy, and the scaling is
the usual coordinate scaling of the density introduced by Levy and
Perdew\cite{LP85} for the GS problem.
Note that this thermal connection formula uses only information between the desired
temperature and higher ones, allowing approximations that begin from the high-temperature
end instead of the low-temperature end\cite{PB16}.  
A second set of formulas give the many relations
among the different correlation energy components (total, potential, and kentropic).
These are very important in ground-state DFT\cite{FTB00,PEB96} for understanding
the origins of different physical
contributions to the correlation free energy and have guided the construction of many approximations.

{\bf Entropy:}
Lastly for this section, we look at a new set of exact conditions for the electronic
entropy as a functional of the density\cite{BSGP16}.  The most important is that the 
universal functional can be written solely in terms of a temperature integral
over entropy, such as
\ben
F\t[\n]=F^0[\n]-\int_0^\tau d\tau'\, S\tp[n],~~~~
A\xc\t[\n]=E\xc[\n]-\int_0^\tau d\tau'\, S\tp\xc[\n],
\label{thent}
\een
i.e., the universal contribution to the free energy functional is
a simple integral over the electronic entropy, and the second shows
that all thermal corrections to the XC  free energy are given by
an integral over the XC entropy.
These formulas have no analog in ground-state DFT.  They also lead to fundamental
inequalities on the various thermal derivatives of both interacting and 
KS quantities.   Such conditions have long been known for the uniform
gas\cite{I82}, but our results are their generalization to inhomogeneous systems.
Analogs are also easily derived from statistical mechanics, but 
again, the tricky part is to deduce their behavior as functionals of the 
density rather than the external potential.  This is why, for example,
all derivatives are total with respect to temperature.  The particle number
is fixed by the density, so temperature is the sole remaining variable.
Our work uses the formalism and methods of ground-state DFT, generalized to finite
temperature, but the same results can also be extracted in the language
of statistical mechanics\cite{DT11,DT16}.

{\bf Tiny violations:}
A minor illustration of the relevance of these conditions is that we found
that a recent parameterization of the thermal XC free energy of the 
uniform gas\cite{KSDT14}
violates one of our conditions for low densities\cite{BSGP16}.  This violation is slight,
and unlikely to ever influence the results of any thLDA calculation.
Nonetheless, it is always better to build parameterizations that satisfy known
conditions, so that the corresponding approximate calculations are guaranteed
to satisfy such conditions\cite{SGVB15,DGSM16}.

{\bf Zero-temperature approximations:}
This work also showed that any ZTA 
calculation automatically satisfies most
of our conditions, whereas the inclusion of thermal XC corrections risks
violating them for specific systems.   
For example, all approximations in Fig. \ref{hubfig} (discussed below) are guaranteed to satisfy
these conditions.
But practical calculations 
including approximate thermal XC corrections should be checked
for possible violations of exact conditions in the future.

\subsection{Exact calculations on a simple model system}
\label{simp}

{\bf Importance for ground-state DFT:}
A crucial step in all DFT development is the exact solution of simple systems
and the test of approximate functionals against exact quantities.  
There are large databases of molecular properties, based either on highly reliable
and accurate experimental measurements, or on far more accurate quantum chemical
calculations\cite{SGB97,ZKP98,KPB99,TS09}.  But these databases usually contain
at most a few numbers per system,
such as the atomization energy and bond length(s).  A substantially more sophisticated
test occurs when a highly accurate calculation is performed such as QMC\cite{CKB79,NU99} or DMRG\cite{W92}, 
and an inversion of the
KS equations\cite{SV09}, so that essentially exact KS potentials, eigenvalues, orbitals, etc, can all be deduced.
This is a much more powerful test of a DFT approximation, and usually provides detailed
insight into its limitations.  The QMC calculations of Umrigar and collaborators\cite{UG94,
FUT94,HU97}
and their high impact, testify to this fact.

{\bf Difficulty of exact thermal calculations:}
Although almost all practical calculations of WDM are in a condensed phase (with 
hot ions), almost all high level inversions yielding exact KS quantities are for
atoms or small molecules.   But even for such systems, it is difficult to
imagine accurate inversions at finite temperature, as only the average particle number
is fixed, and all possible particle numbers must be considered.

\begin{figure}
\centering
\includegraphics[width=1\textwidth]{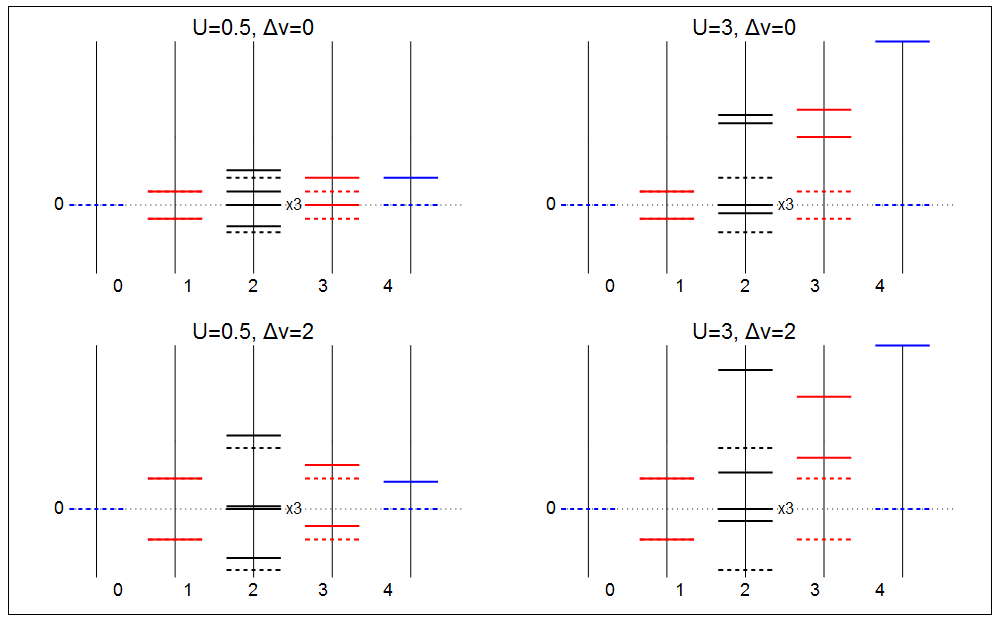}
\caption{
Energy spectrum for the Fock space of the
asymmetric Hubbard dimer at various parameters ranging from 
weakly to strongly correlated. The x-axis is labeled by the number of particles.
The dashed lines correspond to the non-interacting (tight-binding) symmetric case.
The labels in the figures denote degeneracy.  The triplet for $N=2$ is always at $E=0$
due to symmetry.   
The top left is symmetric and weakly correlated, so the spectral lines are close to
the dashed ones.  The top right is symmetric but strongly correlated, and the energies
for $N=2$ are substantially raised.  We also see pairs of levels pushed together.  The
Hubbard bands of the infinite chain roughly run between these levels.
In the lower left panel, we turn on asymmetry, and show that it lessens the effects
of $U$ shown in the upper panels.
 }
\label{spectrum}
\end{figure}

{\bf Exact calculations for Hubbard dimer:}
Exact calculations are only possible for this model because the Hilbert space is severely truncated 
which allows us to compute all energies analytically (see Fig. \ref{spectrum} for complete
diagram of the energy spectrum). However, this means the model is not
even a qualitatively realistic representation of very high temperatures (though we choose
parameters such that the ceiling of the Hilbert space does not effect results).
But we are able to do the inversion exactly,
and so extract all the different contributions to X and C as a function of both 
$\tau$ and $U$.  These are the first exact inversions
of an interacting system 
at finite temperature.  They show us the structure of the underlying functionals, but
cannot tell us which approximations will be accurate.  For example, there is no
real analog of LDA for this system (although BALDA\cite{LSOC03} somewhat plays this role).

\begin{figure}
\centering
\includegraphics[width=.8\textwidth]{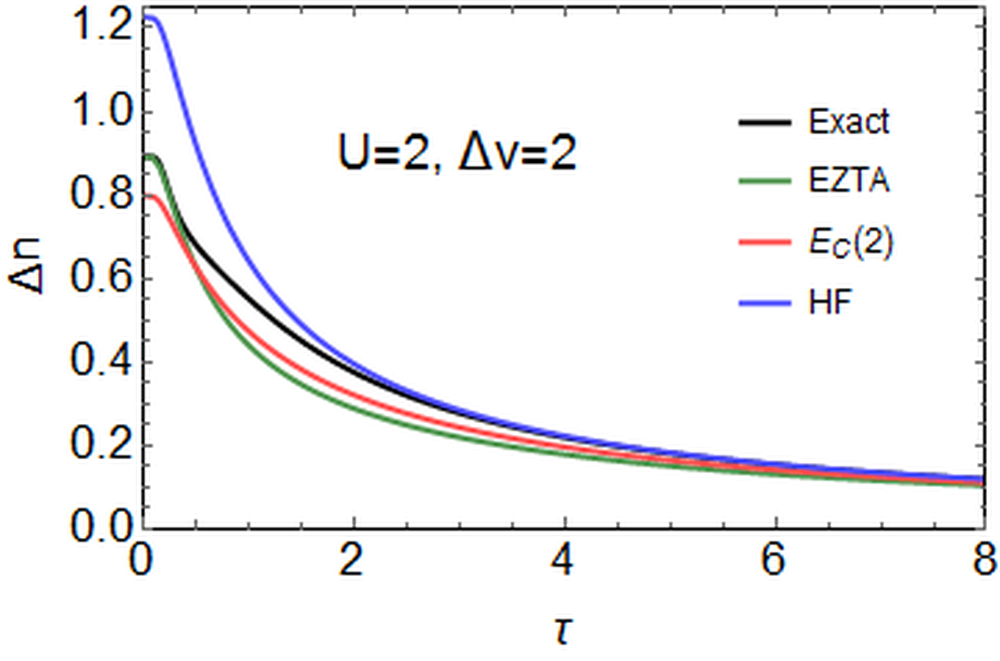}
\caption{
Difference in on-site densities as a function
of temperature for an asymmetric Hubbard dimer with
$U=2$ and site-potential difference $2$ (in units where the hopping parameter
$t=1/2$).  The approximate calculations are all MKS-DFT-ZTA equilibrium
calculations where HF denotes Hartree-Fock, $E\c{(2)}$ includes the
leading correlation correction to HF in powers of the interaction,
while EZTA
denotes using the exact ground-state XC functional\cite{CFSB15}.  }
\label{hubfig}
\end{figure}

{\bf Paradox:}
To see why such simplistic calculations are important,
consider the bottom panel of Fig. \ref{hubfig}.  The black line shows the exact density difference in the
dimer versus temperature for moderate correlation and asymmetry.
The blue curve is a Hartree-Fock calculation, while the red curve adds in the high-density
limit of GS correlation.  Finally, EZTA in green uses the exact
GS functional (i.e. the best possible ZTA),
which we had already found in Ref. \cite{CFSB15},
in the
MKS equations, which therefore is the best possible calculation that
ignores thermal XC contributions.
By construction, this becomes exact in the zero-temperature limit.
But, to our surprise, we
found that the relative error in the free energy and density
vanishes in the {\em high} temperature limit.  In fact, as temperature increases, the fractional
errors at first increase, and then start to lessen.  

\begin{figure}[htb]
\centering
\includegraphics[width=.8\textwidth]{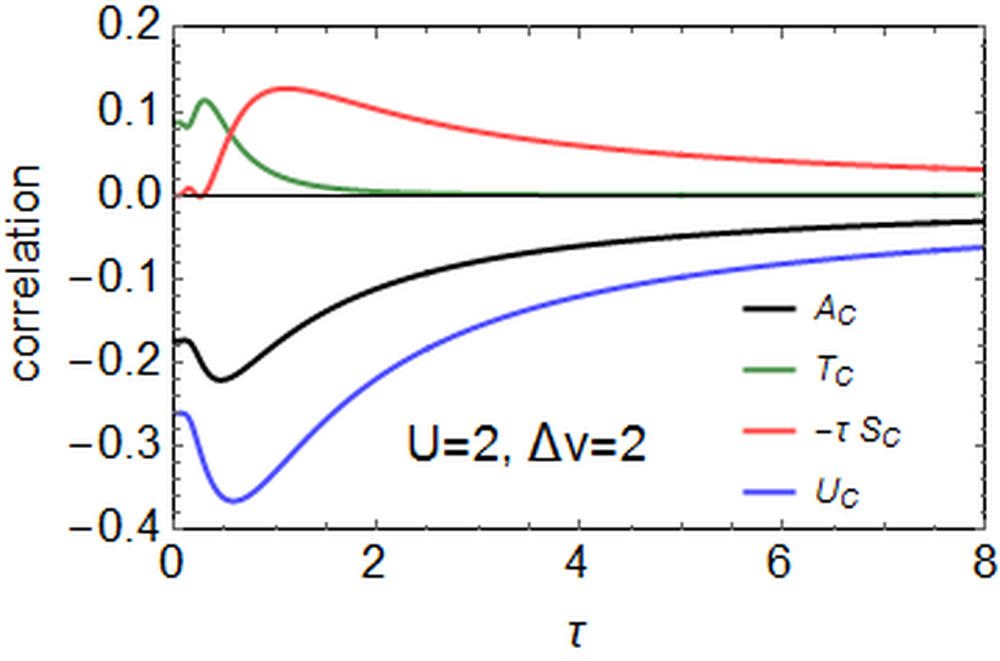}
\caption{
Correlation energy components as a function of temperature. All inequalities proven in Ref.
\cite{PPFS11} are satisfied. This figure shows that the naive assumption that $A\c\t$ is bounded
by $E\c = A\c(\tau=0)$ is not true. Fig. 1 of Ref. \cite{SPB16} shows that the total free
energy increases in magnitude 
as temperature increases, making the XC contribution relatively less
important.  Thus ZTA (or any reasonably bounded approximation) will yield
relatively exact free energies, densities, and KS orbitals, in the
limit of high temperatures.
}
\label{hubXC}
\end{figure}

{\bf Resolution of paradox:}
How can this counterintuitive result be right?  The answer is very simple.  For high
temperatures, the XC components of the energy remain finite (and actually shrink, see Fig. \ref{hubXC}),
while the KS contributions grow, at least in this simple case.  Thus {\em any}
XC approximation will produce the same effect.  This is why
all the approximations merge onto the exact line in Fig. \ref{hubfig}
for sufficiently high temperatures.   Of course, there can
still be a significant absolute error in the free energy which might have important
effects on quantities of interest.  But the principle is clear:  EZTA becomes
relatively exact in both the low- and {\em high}-temperature limits.  This is also
trivially true for the uniform gas, once the (infinite) Hartree energy is included,
and we suspect it to be true for all systems.

{\bf Relevance for response functions:}
The increasing accuracy of the density with temperature has very important implications for calculations of conductivity.  This means
that the error in the KS orbitals, used in the construction of KS conductivities,
starts to decrease beyond some temperature.  This is true for {\em any} GS approximation
for XC (within reason).  All these conclusions may explain the tremendous success so
far gotten by ignoring the thermal XC effects, especially for conductivities.

\subsection{Beyond equilibrium: Linear response thermal time-dependent DFT}

{\bf Zero temperature:}
There are many applications in WDM where the system is perturbed
away from equilibrium.  At zero temperature, the standard approach to such
problems is to apply TDDFT or
many-body non-equilibrium Green's function methods\cite{SV13}.
TDDFT in particular\cite{RG84} can handle
both strong perturbations, such as atoms and molecules in intense laser fields,
and weak perturbations, where the linear response formulation yields excitation energies and oscillator
strengths\cite{BWG05}.  

{\bf General case too difficult:}
Unfortunately, the situation is very complicated if the perturbation is strong,
as then a non-equilibrium treatment is needed.  Theories in which the temperature
is held fixed do not apply.   This is the situation for example in calculations
of stopping power\cite{GSK96}.   There
are many fine attempts to overcome these difficulties
under a variety of practically useful conditions\cite{RTKC16}, but we have not seen a way
to construct a general DFT treatment of such problems.

{\bf Linear response:}
For a finite system (which
has to be very carefully defined in the thermal case), we  proved
a limited theorem for the linear density response to a time-dependent external 
field\cite{PGB16}.
This proof allows for (finite numbers of) degeneracies in the excited states,
but not in the equilibrium state.  Armed with such a theorem, all the usual
XC response properties, such as the XC kernel, can be defined at finite temperature.
Combined with our thermal connection formula, we have the finite-temperature
generalization of the Gross-Kohn response equation\cite{GK85}:
\ben
\chi\t(12)=\chi\s\t(12) + \int d3d4\, \chi\s\t(13) f\Hxc\t(34) \chi\t(42),
\label{thGK}
\een
where $1$ denotes the coordinates $\br,t$, and 2 another pair\cite{KBP94},
$\chi\t(12)$ is the density-density response function at temperature $\tau$,
$\chi\s$ its KS counterpart, and $f\Hxc\t(12)$ the thermal Hartree-XC kernel.
This becomes the Random Phase Approximation (RPA) when $f\xc =0$.
Insertion of this into the thermal connection formula yields an RPA-type
equation for the XC free energy\cite{PGB16}:
\ben
A\xc\t[\n]=
\lim_{\tau''\to\infty}\frac{\tau}{2} \int_\tau^{\tau''} \frac{d\tau'}{\tau'^2}\,
\int_0^\infty \frac{d\omega}{2\pi}\coth{\bigg(\frac{\omega}{2\tau}\bigg)}
\int d\br\int d\br'
 \frac{\Im\chi^{\tau'}[\n\g](\br,\br',\omega)}
{|\br-\br'|}.
\label{AcFD}
\een
If XC contributions to the kernel are neglected, this becomes the long-known
random-phase approximation to the XC free energy, albeit using the KS
orbitals.
Since random-phase approximation calculations have become standard within the GS 
materials world\cite{SHSG10},
there is little additional computational demand in performing them at moderate finite
temperatures.  
Inclusion of any approximate treatment of the XC kernel yields an 
entirely novel approach to XC approximations
for equilibrium thermal DFT.  In particular, one can consider making a uniform
approximation  in both space and time, and also decide whether or not to include
thermal corrections in an approximate kernel.
All such treatments can be first tested on a uniform gas, for which the XC free
energy is accurately known from QMC calculations\cite{DGSM16}.

\section{Recent applications of DFT in WDM}
\label{apps}

{\bf Planetary science:}
The materials of interest in WDM research span the periodic table.
Accurate thermal conductivities for inertial confinement fusion fuel materials
such as deuterium and tritium 
are needed to find the calculated total neutron yield in fusion science
(the simulated mixing between the fuel and coating on ICF capsules is
very sensitive to thermal conductivities)\cite{HCBK14}.
Accurate iron thermal conductivities are used to determine whether
the conventional model for how the Earth's core developed is
valid because heat flux contributes to the Earth's geomagnetic field.
Differential heating experiments at ALS, LCLS, Omega, and Titan facilities are
all done slightly differently (heating via optical lasers, XFEL, x-rays, or proton heating),
to fit hydrodynamics models because of their high accuracy and because other
approaches (SESAME, Purgatorio, and Lee-More) all differ\cite{BJKR78,LM84,WSSI06}.
Our work suggesting that ignoring XC thermal corrections nevertheless yields
accurate KS eigenstates and eigenvalues\cite{SPB16} helps explain why conductivities
can be accurate in these calculations.

Much WDM research is motivated by the desire to understand planetary interiors.
The Juno mission is measuring Jupiter's gravitational field extremely
accurately, constraining theories of its interior\cite{MGF16}, while Kepler has
shown that many notions of planetary formation must be rethought with
our new data on extra-solar planets\cite{C16}.  But there is limited understanding
of whether initial planetary protocores remain stable during accretion
or if they dissolve into outer metallic hydrogen layers.  Recent
DFT-MD calculations show that MgO is surprisingly soluble in hydrogen under
these conditions\cite{WM12}.
Similarly, the moon is thought to have formed in an enormous impact, but such
a scenario depends crucially on the equation of state of MgO under extreme
conditions.  Recent DFT calculations and Z-machine experiments have
nailed this EOS more accurately than before, and far better than unreliable
extrapolations from more mundane conditions\cite{RSLD15}.

{\bf Alternate methods:}
Path integral Monte Carlo is an excellent tool for studying WDM, and has been
recently extended beyond small atoms to include water and carbon,
and has recently been shown to match reasonably
well with DFT calculations at lower temperatures\cite{DM12}, validating both.
Meanwhile, DFT calculations have predicted new
superionic phases of H$_2$O, under conditions
relevant to Uranus and Neptune interiors\cite{WWM13}.

{\bf DFT failure:}
A less successful application of DFT in WDM is to the liquid insulator to liquid metal
transition in dense D$_2$, at about 1000K and 300 GPa.  DFT calculations
with several different functionals yield very different results, none of
which are in satisfactory agreement with experiment.  The interpretation
also depends on the accuracy of the conductivity from the DFT calculations.
This system remains a challenge to WDM simulations.

{\bf X-ray Thompson scattering:}
Some of the most exciting recent experiments have been from the LINAC
at SLAC, allowing X-ray Thompson scattering (XRTS) measurements of shocked
materials.  These include the first highly resolved measurements of the plasmon spectrum in an
ultrafast heated solid\cite{SGLC15}.
Ref. \cite{DDRF16} gives x-ray scattering results from plasmons in
dynamically compressed deuterium, from which one can deduce the
ionization state as a function of compression.  Ionization begins
at about the pressure that DFT-MD calculations show molecular dissociation.
In a completely different material, X-ray diffraction showed diamond
formation on nanosecond timescales, caused by shock compression to about
200 GPa\cite{KRGG16}.   This helps explain why the lonsdaleite crystal
structure occurs naturally close to meteor impacts.

XRTS has been performed on a variety of materials including Be, Li, C, CH shells, and Al.
Most experiments probe the electron dynamic structure factor, which is
decomposed via the somewhat ad-hoc Chihara decomposition into bound, loosely
bound, and free electrons\cite{BSDH16}.  But by running TDDFT at finite temperatures,
one directly calculates the densities,
and can then test the accuracy of Chihara for determining the ionization state.
The results of Ref. \cite{PGB16} are already being used to justify thTDDFT calculations
such as Ref. \cite{BSDH16}.

\section{Relation of thermal DFT to quantum chemistry}

At first glance, it would appear that warm dense matter has little or nothing
to do with chemistry.  In fact, this is not true, it is simply chemistry in 
an exciting new regime with which we are relatively unfamiliar.

To see this, we first note that the plasma physicists who usually study WDM think 
in terms of average properties of their systems, such as mean densities and numbers
of electrons ionized.  They are familiar with density functional methods, but traditionally
only at the level of the LDA.  Successes with such an approximation
are often attributed to systems being somehow `locally uniform'.

But the success of DFT methods in chemistry can be directly correlated with the arrival
of the GGA and hybrids of it with Hartree-Fock.  These
approximations were tested on the G2 data set, and shown to yield much better energetics
than LDA, because the G2 data set had already been carefully constructed and benchmarked,
using both quantum chemical methods and experimental information\cite{CRRP97,CRRP98,PHMK05}.  This vote of confidence
led to their widespread adoption in many branches of chemistry, and also led to the
confidence that GGAs were better than LDA for many materials problems.

It is the same GGAs, used in MD simulations, that have led to the revolution in WDM
simulations over the past decade or so\cite{LBKC00}.  The improved accuracy due to GGAs implies
that the details of the electronic structure matter, and that these systems are in
no way locally uniform.  In fact, in many cases, there are large evanescent regions
of the HOMO, just as in gas-phase molecules.  The KS system is ideal for
computing this, and GGAs and hybrid account for the energetic consequences.
So the very success of DFT-MD for WDM implies that the detailed chemistry is vital, even
if it is happening within simulations of extended systems under high temperature and
pressure.

The recent work in our group is almost entirely focused on bringing GGA-level XC technology
to the WDM field.   Being able to distinguish among different components of the
correlation energy, and switch from one to another, is a crucial part of the exact
conditions that were used to construct GGAs\cite{PB16}.  The adiabatic connection formula
is often invoked in modern DFT research to understand both exact DFT and approximations,
and its recasting as a temperature integral should prove useful in the search for accurate
thermal XC approximations.   

On the other hand, ground-state DFT has benefited enormously from testing on benchmark data\cite{SGB97,ES99}.
But for thermal effects, even a simple H atom is difficult, as one must include sums over
all possible particle numbers in the partition function.   The
asymmetric Hubbard dimer is the simplest imaginable exactly-solvable model, and can be
considered a model for H$_2$ in a minimal basis.  While the truncated Hilbert space makes
it unrealistic at higher temperatures, it also makes it practical to solve exactly.
Thus our calculations on this model demonstrate the behavior of correlation at finite
temperatures in one simple case.  Unfortunately, due to the lack of a continuum, this
cannot be used to check the performance of LDA or GGA.  

Thermodynamics tells us simple relations between entropy and free energy and other quantities.
But it requires very careful reasoning to deduce the corresponding relations among density
functionals, as the density must be held fixed, not the external potential.  Our relations
between entropy and the universal part of the Mermin functional show this, and subtraction
of the corresponding KS contributions yields crucial relations among correlation contributions.
They also yield simple inequalities that are not automatically satisfied once thermal XC
contributions are approximated.

Lastly, the recent proof of TDDFT for finite temperatures within linear response justifies the
extraction of conductivities within the Kubo response formalism from KS orbitals and energies.
It also shows that the random-phase approximation, which is now routinely calculated
for inhomogeneous systems in many codes in both quantum chemistry and materials
science\cite{KF96,F01,F08,EYF10,EBF12,F16},
might be an excellent starting point for more accurate approximations to the
XC thermal corrections, using approximations to the temperature-dependent XC kernel.

Finally, our recent work explains how ignoring thermal XC effects, which is usually done
in practical DFT WDM calculations, might not be as poor an approximation as it first
appears.  Calculations on the Hubbard dimer show that XC effects become relatively
less important as the temperature increases.  Thus the errors in the self-consistent
density and orbitals caused by any approximation
to the XC lessen with increasing temperature, so that calculations of the KS
conductance should be more accurate as temperature increases (somewhat counterintuitively).

To summarize, the success of modern density functional approximations in WDM simulations
strongly implies the importance of chemical phenomena in such simulations, and the
need to accurately approximate the energetics.

\section{Conclusion}

Thermal density functional theory is an increasingly utilized tool for calculations of hot systems such as warm dense matter. These WDM systems include inertial confinement fusion, planetary interiors, and shock experiments. 
There have been many recent developments ranging from exact conditions, improved understanding,
and extensions beyond equilibrium with more foreseeable, and exciting, improvements on the horizon. 
These steps forward set up the foundation for further future success of thermal DFT 
in the years to come.

\begin{acknowledgement}
The authors acknowledge support from the US Department of Energy (DOE), Office of Science, Basic Energy Sciences under Award No. DE-FG02-08ER46496. 
J.C.S. acknowledges support through the NSF Graduate Research fellowship program under Award No. DGE-1321846. 
\end{acknowledgement}


\printbibliography

\end{document}